\documentclass{svproc}

\usepackage[T1]{fontenc}
\usepackage{color}

\usepackage{graphicx}  
\usepackage{subfigure}

\usepackage{amsmath}
\usepackage{amssymb}

\usepackage{url} 

\let\epsilon \varepsilon
\begin{document}
\mainmatter              
\title{An open and parallel multiresolution  framework using block-based adaptive grids}
\titlerunning{A new multiresolution code using block-structured grids}  
\author{Mario Sroka\inst{1} \and Thomas Engels\inst{2} \and Philipp Krah\inst{1} \and Sophie Mutzel\inst{1} \and Kai Schneider\inst{3} \and Julius Reiss\inst{1}}
\authorrunning{Sroka, Engels, Krah, Mutzel, Schneider, Reiss} 
%
\tocauthor{Sroka, Engels, Krah, Mutzel, Schneider, Reiss}

\institute{Technische Universit\"at Berlin, M\"uller-Breslau-Strasse 15, 10623 Berlin, Germany,\\
\and
\'{E}cole normale sup\'{e}rieure,
LMD (UMR 8539), 24, Rue Lhomond, 75231 Paris Cedex 05, France,\\
\email{thomas.engels@ens.fr}
\and
Aix--Marseille Universit\'{e}, CNRS, Centrale Marseille, I2M UMR 7373,
39 rue Joliot-Curie, 13451 Marseille Cedex 20 France \\
}

\maketitle              

\begin{abstract}
A numerical approach for solving evolutionary partial differential equations in two and three space dimensions on block-based adaptive grids is presented. 
The  numerical discretization is based on high-order, central finite-differences and explicit time integration. 
Grid refinement and coarsening are triggered by multiresolution analysis, i.e. thresholding of wavelet coefficients, which allow controlling the precision of the adaptive 
approximation of the solution with respect to uniform grid computations. 
The implementation of the scheme is fully parallel using MPI with a hybrid data structure. 
Load balancing relies on space filling curves techniques. Validation tests for 2D advection equations allow to assess the precision and performance of the developed code. 
Computations of the compressible Navier-Stokes equations for a temporally developing 2D mixing layer illustrate the properties of the code for nonlinear multi-scale problems.
The code is open source. 
\end{abstract}

\section{Introduction} \label{intro}

For many applications in computational fluid dynamics, adaptive grids are more advantageous than uniform grids, because computational efforts are put at locations required by the solution.
Since small-scale flow structures may travel, emerge and disappear, the required local resolution is time-dependent. 
Therefore dynamic gridding, which tracks the evolution of the solution, is more efficient than static grids. 
However, suitable grid adaptation techniques are necessary to dynamically track the solution. 
These techniques can increase the computational cost, therefore their efficiency is problem dependent and related to the sparsity of the adaptive grid.

Examples where adaptivity is beneficial are reactive flows with localized flame fronts, detonations and shock waves  \cite{BengoecheaGrayMoeckPaschereitSesterhenn2018,Roussel2015}, coherent structures in turbulence \cite{SchneiderVasilyev2010} and
flapping insect flight \cite{EngelsKolomenskiySchneiderSesterhenn2016}. 
For the latter the time-varying geometry generates localized turbulent flow structures.
These applications motivate and trigger the development of a novel multiresolution framework,
which can be used for many mixed parabolic/hyperbolic partial differential equations (PDE).

The idea of adaptivity is to refine the grid where required and to coarsen it where possible, while controlling the precision of the solution.

Such approaches have a long tradition and can be traced back to the late seventies \cite{Brandt1977}. 
Adaptive mesh refinement and multiresolution concepts developed by Berger et al. \cite{BergerOliger1984} and Harten \cite{Harten1993,Harten1996}, respectively, are meanwhile widely used  for large scale computations (e.g. \cite{Deiterding2011,Mueller2003,DominguesGomesRousselSchneider2011}). 

Berger  suggested  a flexible refinement strategy by overlaying different grids of various orientation and size, in the following referred to as \emph{adaptive mesh refinement} (AMR).
Harten instead discusses a mathematical more rigorous wavelet based method, termed \emph{multiresolution} (MR). 
For AMR methods, the decision where to adapt the grid is based on error indicators, such as  gradients of the solution or derived quantities. 
In contrast  in MR, the   multiresolution transform allows efficient compression of data fields by thresholding detail coefficients.
This multiresolution transform  is equivalent to biorthogonal wavelets, see e.g. \cite{Harten1996}.
An important feature of MR is the reliable error estimator of the solution on the adaptive grid, as the error introduced by removing grid points can be directly controlled.

In wavelet-based approaches the governing equations are discretized,  either by  using wavelets in a Galerkin or collocation approach 
\cite{SchneiderVasilyev2010}, or using a classical discretization, e.g. finite volumes or differences, where the grid is adapted locally using MR analysis \cite{BramkampLambyMueller2004,DominguesGomesRousselSchneider2011}. 

MR methods typically keep only the information which is  dictated by a threshold criterion, which is refereed to as  sparse point representation
(SPR), introduced in \cite{Holmstroem1999}. 
		AMR methods  often utilize blocks and refine complete areas, by which the maximal sparsity is typically abandoned in favor of a simpler code  structure. 
		An example of this approach is the AMROC code \cite{DeiterdingDominguesGomesRousselSchneider2009}, where blocks of arbitrary size and shape are refined. 
		A detailed  comparison of MR with AMR techniques has been carried out in \cite{DeiterdingDominguesGomesSchneider2016}.

For practical applications both the data compression and the speed-up of the calculation are crucial.
The latter is reduced by the computational overhead to handle the adaptive grid and corresponding datastructures.
This effort differs substantially between different approaches \cite{Mueller2009}. 
It can be reduced by refining complete blocks, thereby reducing the elements to manage, and by exploiting simple grid structures.    

A MR method using a quad- or octtree representation to simplify the grid structure is reported, e.g., in \cite{DominguesGomesDiaz2003,DominguesGomesRousselSchneider2011} and has later also been used in \cite{RossinelliHejazialhosseiniSpampinatoKoumoutsakos2011}.

For detailed reviews on the subject of multiresolution methods we refer the reader to \cite{CoquelMadayMullerPostelTran2009,SchneiderVasilyev2010,Mueller2003,SchneiderVasilyev2010,DominguesGomesRousselSchneider2011}. 
Implementation issues have been discussed in \cite{BrixMelianMullerBachmann2011}.

Our aim is to provide a multiresolution framework, which can be easily adapted to different two- and three-dimensional simulations encountered in CFD, and which can be efficiently used on fully parallel machines.

To this end the chosen framework is block based, with nested blocks on quad- or octree grids.
The individual  blocks  define structured grids with a fixed number of points.
Refinement and coarsening are controlled by a threshold criterion applied to the wavelet coefficients. 
The software, termed ``wavelet adaptive block-based solver for interactions
with turbulence'' (\verb+WABBIT+),  is open-source and freely available\footnote{Available on https://github.com/adaptive-cfd/WABBIT}
in order to maximize its utility for the scientific community and for reproducible science. 


The purpose
of this paper is to introduce the code, present its main features
and explain structural and implementation details. It is organized
as follows. In section \ref{code} we give an overview of implementation
and structure details. Numerics will only be shortly described, but
special issues of our data structure, interpolation, and the \verb+MPI+
coding will be explained in detail. Section \ref{swirl} considers
a classical validation test case, including a discussion on the adaptivity and convergence
order of \verb+WABBIT+. In section \ref{dsl} we present computations for a temporally developing double shear layer, governed by the compressible Navier-Stokes equations. Section \ref{con} draws conclusions and gives perspectives
for future work.

\section{Code structure} \label{code}
In this section we present a detailed description
of the data and code structure. One of the main concepts in WABBIT
is the encapsulation and separation of the set of PDE from the rest
of the code, thus the PDE implementation is not significantly different
from that in a single domain code and can easily be exchanged. The
code solves evolutionary PDE of the type $\partial_{t}\phi=N(\phi)$.
The spatial part $ N(\phi)$ is referred  to as \emph{right hand side} in this report. 
A primary directive for the code is its ``explicit simplicity'', which
means avoiding complex programming structures to improve maintainability.
WABBIT is written in Fortran~95 and aims at reaching high performance
on massively parallel machines with distributed memory architecture.
We use the MPI library to parallelize all subroutines, while parallel
I/O is handled through the HDF5 library.

\subsection{Multiresolution algorithm}

The main structure of the code is defined by the
multiresolution algorithm. After the initialization phase, the general
process to advance the numerical solution $\varphi\left(t^{n},x\right)$
on the grid $\mathcal{G}^{n}$ to the new time level $t^{n+1}$ can
be outlined as follows. 
\begin{enumerate}
\item Refinement. We assume that the grid $\mathcal{G}^{n}$
is sufficient to adequately represent the solution $\varphi\left(t^{n},x\right)$,
but we cannot suppose this will be true at the new time level. Non-linearities
may create scales that cannot be resolved on $\mathcal{G}^{n}$, and transport can advect existing fine structures.
Therefore, we have to extend $\mathcal{G}^{n}$ to $\mathcal{\widetilde{G}}^{n}$
by adding a ``safety zone'' \cite{SchneiderVasilyev2010} to ensure that
the new solution $\varphi\left(t^{n+1},x\right)$ can be represented
on $\mathcal{\widetilde{G}}^{n}$. To this end, all blocks are refined
by one level, which ensures that quadratic non-linearities cannot
produce unresolved scales.
\item Evolution. On the new grid $\mathcal{\widetilde{G}}^{n}$,
we first synchronize the layer of ghost nodes (section \ref{data_synch})
and then solve the PDE using finite differences and explicit time-marching
methods. 
\item Coarsening. We now have the new solution $\varphi(t^{n+1},x)$
on the grid $\mathcal{\widetilde{G}}^{n}$. The grid $\mathcal{\widetilde{G}}^{n}$
is a worst-case scenario and guarantees resolving $\varphi(t^{n+1},x)$
using \emph{a priori}  knowledge
on the non-linearity. It can now be coarsened to obtain the new grid $\mathcal{G}^{n+1}$,
removing, in part, blocks created during the refinement stage. Section
\ref{subsec:Refinement--coarsening} explains this process in more
detail.
\item Load balancing. The remaining blocks are, if necessary,
redistributed among MPI processes using a space-filling curve \cite{Zumbusch2003}, 
such that all processes compute approximately the same number of blocks.
The space-filling curve allows preservation of locality and reduces
interprocessor communication cost.
\end{enumerate}

\subsection{Block- and Grid Definition}

\paragraph{Block definition.}

The decomposition of the computational domain builds on blocks as
smallest elements, as used for example in \cite{DominguesGomesDiaz2003}. The
approach thus builds on a hybrid datastructure, combining the advantages
of structured and unstructured data types. The structured blocks have
a high CPU caching efficiency. Using blocks instead of single points
reduces neighbor search operations. A drawback of the block based
approach is the reduced compression rate.

\begin{figure}
\begin{centering}
\includegraphics{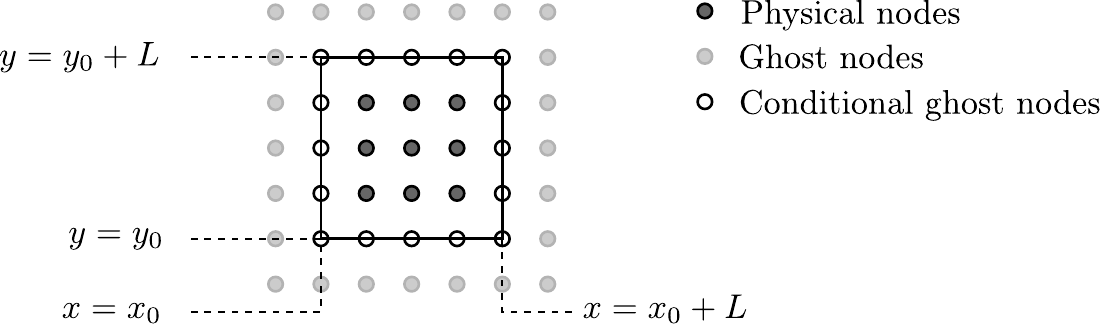} 
\par\end{centering}
\caption{Definition of a block with $B_{s}=5$ and $n_{g}=1$. \label{fig:Definition-of-a}}
\end{figure}

A block is illustrated in Fig. \ref{fig:Definition-of-a}. Its definition
(in 2D) is 
\[
\mathcal{B}^{\ell}=\left\{ \underline{x}=\underline{x}_{0}+\left(i\cdot\Delta x^{\ell},j\cdot\Delta x^{\ell}\right)^{T},\;0\leq i,j\leq B_{s}\right\} 
\]
where $\underline{x}_{0}$ is the blocks origin, $\Delta x^{\ell}=2^{-\ell}L/(B_{s}-1)$
is the lattice spacing at level $\ell$, 
and $L$ the size of the entire computational domain. The mesh level
encodes the refinement from 1 as coarsest to the user defined value
$J_{\mathrm{max}}$ as finest. Blocks have $B_{s}$ points in each
direction, where $B_{s}$ is odd, which is a requirement of the grid
definition we use. We add a layer of $n_{g}$ ghost
points that are synchronized with neighboring blocks (see section
\ref{data_synch}). The first layer of physical points is called
conditional ghost nodes, and they are defined as follows:
\begin{enumerate}
\item If the adjacent block is on the same level, then the conditional ghost
nodes are part of both blocks and thus redundant in memory; their
values are identical. 
\item If the levels differ, the conditional ghost nodes belong to the block
on the finer level, i.e., their values will be overwritten by those
on the finer block. 
\end{enumerate}

\paragraph{Grid definition.}

\begin{figure}[htb]
\begin{centering}
\includegraphics{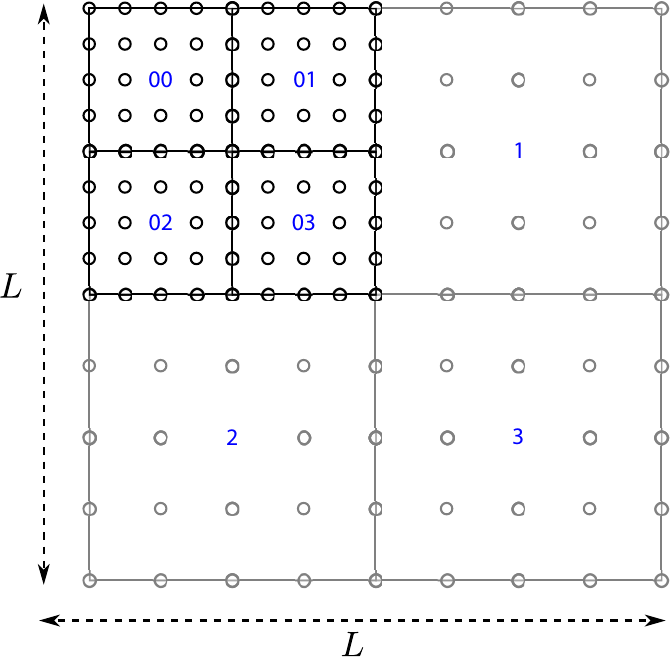} 
\par\end{centering}
\caption{Example grid with $N_{b}=7$ blocks. Three blocks
on mesh level 1 (gray) and four on level 2 (black),
together with their treecodes. Note that the mesh level is equal to
the length of the treecode. Points at the coarse/fine interface belong
to finer blocks.}
\label{fig_cod_001} 
\end{figure}

A complete grid consisting of $N_{b}=7$ blocks is shown in Fig. \ref{fig_cod_001}.
We force the grid to be graded, i.e., we limit the
maximum level difference between two blocks to one. Blocks are addressed
by a quadtree-code (or an octtree in 3D), as introduced in
\cite{Gargantini1982}, and also shown in Fig. \ref{fig_cod_001}.
Each digit of the treecode represents one mesh level, thus its length
indicates the level $\ell$ of the block. If a block is coarsened,
the last digit is removed, while for refinement refinement, one digit is added. The
function of the treecode is to allow quick neighbor search, which
is essential for high performance. For a given treecode the adjacent
treecodes can easily be calculated \cite{Gargantini1982}. A list
of the treecodes of all existing blocks allows us to find the data
of the neighboring block, see section \ref{subsec:Data-Structure}.
To ensure unique and invertible neighbor relations, we define them
not only containing the direction but also 
 encode if a block covers only part a  border. This situation occurs if two neighboring blocks differ in level.  
We also account for diagonal neighborhoods. 
In two space dimensions  16 different relations defined (74 in 3D). 
This simplifies the ghost nodes synchronization step, since all required information,
the neighbor location and interpolation operation are available. 

\paragraph{Right hand side evaluation.}

The PDE subroutine purely acts on single blocks.
Therefore efficient, single block finite difference schemes can be
used allowing to combine existing codes with the WABBIT framework.
Adapting the block size to the CPU cache offers near optimal performance
on modern hardware. The size of the ghost node layer can be chosen
freely, to match numerical schemes with different stencil sizes. 

\subsection{Refinement / coarsening of blocks.\label{subsec:Refinement--coarsening}}

\begin{figure}
\begin{centering}
\includegraphics{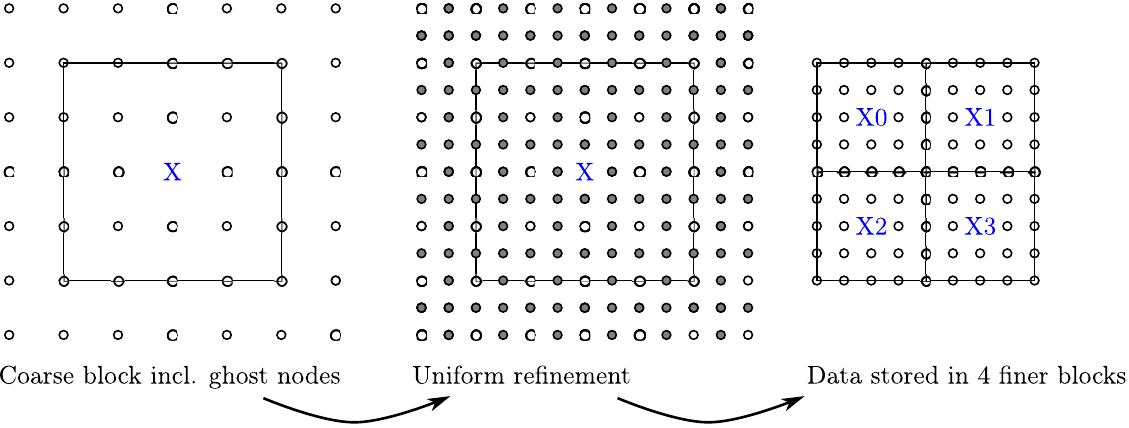} 
\par\end{centering}
\caption{Process of refining block with treecode X. First, the block is upsampled,
including the ghost nodes layer. Then, four new blocks are created,
where one digit is added to the treecode.\label{fig:Process-of-refining}}
\end{figure}

If a block is flagged for refinement by some criteria (see
blow) this refinement is executed as illustrated in Fig. \ref{fig:Process-of-refining}.
The block, with synchronized ghost points, is first uniformly upsampled
by midpoint insertion, i.e., missing values on the grid 
\[
\mathcal{\widetilde{B}}^{\ell}=\left\{ \underline{x}=\underline{x}_{0}+\left(i\cdot\Delta x^{\ell}/2,j\cdot\Delta x^{\ell}/2\right)^{T},\quad-2n_{g}\leq i,j\leq2B_{s}-1+2n_{g}\right\} 
\]
are interpolated (gray points in Fig. \ref{fig:Process-of-refining}
center). In other words, a prediction operator $\mathcal{P}_{\ell\rightarrow\ell+1}$
is applied  \cite{Harten1993}. The data is then distributed to four
new blocks $\mathcal{B}_{i}^{\ell+1}$, where one digit is added to
the treecode, which are created on the \verb+MPI+ process holding
the initial (``mother'') block. The blocks are
nested, i.e. all nodes of a coarser block also exist
in the finer one. The reverse process is coarsening,
where four sister blocks on the same level are merged into one coarser
block by applying the restriction operator $\mathcal{R}_{\ell\rightarrow\ell-1}$,
which simply removes every second point. For coarsening, no ghost
node synchronization is required, but all four blocks need to be gathered
on one \verb+MPI+ rank. 

The refinement operator uses central interpolation schemes. Using
one-sided schemes close to the boundary would not require 
ghost points and would thus reduce the number of communications.
They yield errors only of the order  of the threshold $\varepsilon$.
However, the small, but
non-smooth structures of these errors force very fine meshes, which
can increase the number of blocks. This \emph{fill-up} can lead to
prohibitively expensive calculations.

\paragraph{Computation of detail coefficients.}

The decision whether a block can be coarsened or not is made by calculating
its detail coefficients \cite{SchneiderVasilyev2010}. The are computed by
first applying the restriction operator, followed
by the prediction operator. After this round trip of restriction
and prediction, the original resolution is recovered, but the values
of the data differ slightly. The difference 
\[
\mathcal{D}=\{d(\underline{x})\}=\mathcal{B}^{\ell}-\mathcal{P}_{\ell-1\rightarrow\ell}(\mathcal{R}_{\ell\rightarrow\ell-1}(\mathcal{B}^{\ell}))
\]
is called details. 
If details are small, the field is smooth on the current grid level. Therefore, the details act as indicator for a possible coarsening \cite{Harten1993}. 
Non-zero details are obtained at odd indices only
(gray points in Fig. \ref{fig:Process-of-refining}, center) because
of the nested grid definition and the fact that restriction and prediction
do not change these values. The refinement flag for a block is then
\[
r=\begin{cases}
-1 & \text{if }\left\Vert d(\underline{x})\right\Vert _{\infty}<\varepsilon\\
\phantom{-}0 & \text{otherwise}
\end{cases}
\]
where -1 indicates coarsening and 0 no change. In other words, the
largest detail sets the status of the block. Note, that WABBIT technically
provides the possibility to flag -1 for coarsening, 0 for unaltered
and +1 for refinement, it can hence be used with arbitrary indicators.
Since a block cannot be coarsened if its sister blocks on the same
root do not share the +1 refinement status, WABBIT assigns the -2
status for blocks that can indeed be coarsened, after checking for
completeness and gradedness.

\subsection{Data structure\label{subsec:Data-Structure}}

The data are split into two kinds of data, first,
the field data (the flow fields) required to calculate the PDE 
and, second, the data to administer the block decomposition and the
parallel distribution.

Data which are held only on one specific \verb+MPI+
process are called \textit{heavy data}. This is
the (typically large) field data and the neighbor relations for the
blocks held by the MPI process. The field data (\verb+hvy_block+)
is a five dimensional array where the first three indices describe
the note within a block (3D notation is always used in the code),
the fourth index the index of the physical variables and the last one the
block index identifying it within the MPI process. 

The \textit{light data }(\verb+lgt_block+) are
data which are kept synchronous between all processes.
They  describe the global topology of the adapted grid and change during
the computation. The light data  consist of the block treecode, the block mesh
level and the refinement flag. Additionally, we
encode the \verb+MPI+ process rank $i_{\mathrm{process}}$ and the
block index on this process $j_{\mathrm{block}}$ by the position
$I$ of the data within the \textit{light data} array, $I=(i_{\mathrm{process}}-1)\cdot N_{\mathrm{max}}+j_{\mathrm{block}}$,
where $N_{\mathrm{max}}$ is the maximal number of blocks per process.
The \emph{light data} enable each process to determine the process holding neighboring
blocks, by looking for the index $I$ corresponding to the adjacent
treecode. The number of blocks required during the 
computation is unknown before running the simulation. To avoid time
consuming memory allocation, $N_{\mathrm{max}}$ is typically determined
by the available memory. This sets the index range of the last index
of the heavy data and determines the size of the light data. Hence,
many blocks are typically unused; they are marked by setting the treecode
in the light data list to -1. To accelerate the search within the
\emph{light data}, we keep a second
list of indices holding active entries.

\subsection{Parallel implementation\label{data_synch}}

\paragraph{Data synchronization.}

For parallel computing, an efficient data synchronization strategy
is essential for good performance. There are two different tasks in
WABBIT, namely light and heavy data synchronisation. \textit{Light
data} synchronization is an MPI all-to-all operation,
where we communicate active entries of the light data only.\textit{
Heavy data} synchronization, i.e. filling the ghost
nodes layer of each block, is much more complicated. We
have to balance a small number of MPI calls and a small amount of
communicated data, and additionally we have to ensure that no idle
time occurs due to blocking of a process by a communication in which
this process is not involved. To this end, we use
\verb+MPI+ point-to-point communication, namely non-blocking non-buffered
send/receive calls. To reduce the number of communications, the ghost
point data of all blocks belonging to one process are gathered and
send as one chunk. After the \verb+MPI+ communications, all processes
store received data in the ghost point layers. 

The conditional ghost nodes require special attention during the synchronization.
To ensure that neighboring blocks always have the same values at these nodes, the
redundant nodes are sent, when required, to the neighboring process.
Blocks on higher mesh levels (finer grids) always overwrite the redundant
nodes to neighbors on lower mesh level (coarser grid). It is assumed
that two blocks on the same mesh level never differ at a redundant
node, because any numerical scheme should always produce the same
values.

%
%

\paragraph{Load balancing.}

The external neighborhood consists of ghost nodes, which may be located
on other processes and therefore have to be sent/received in the \textit{heavy
data} synchronization step. Internal ghost nodes can simply be copied
within the process memory, which is much faster than \verb+MPI+ communication.
It is, thus, desired to reduce inter-process neighborhood. We use space
filling curves \cite{Zumbusch2003} to redistribute the blocks among the processes for their good localization. 
The computation of the space filling curve is simple, because we can
use the treecode to calculate the index on the curve. 


\section{Advection test case} \label{swirl}

As a validation case, we now consider a benchmarking problem for the
2D advection equation, $\partial_{t}\varphi+u\cdot\nabla\varphi=0$,
where $\varphi(x,y,t)$ is a scalar and $0\leq x,y<1$. The spatially-periodic
setup considers time-periodic mixing of a Gaussian blob,
\[
\varphi(x,y,0)=e^{-\left((x-c)^{2}+(y-d)^{2}\right)/\beta}
\]
 where $c=0.5$, $d=0.75$ and $\beta=0.01$. The time-dependent velocity
field is given by 
\begin{equation}
u(x,y,t)=\cos\left(\dfrac{\pi t}{t_{a}}\right)\left(\begin{array}{cc}
\sin^{2}(\pi x)\sin(2\pi y)\\
\sin^{2}(\pi y)(-\sin(2\pi x))
\end{array}\right)\label{eqn_swirl_002}
\end{equation}
and swirls the initial distribution, but reverses to the initial state
at $t=t_{a}$. 
The swirling motion produces increasingly fine structures
until $t=t_{a}/2$, where $t_{a}$ controls also the size of structures.
The larger $t_{a}$, the more challenging is the test.

\begin{figure}[htb]
	\begin{center}  
		\includegraphics[width=.31\textwidth,clip=true,viewport= 60 10 360 310]{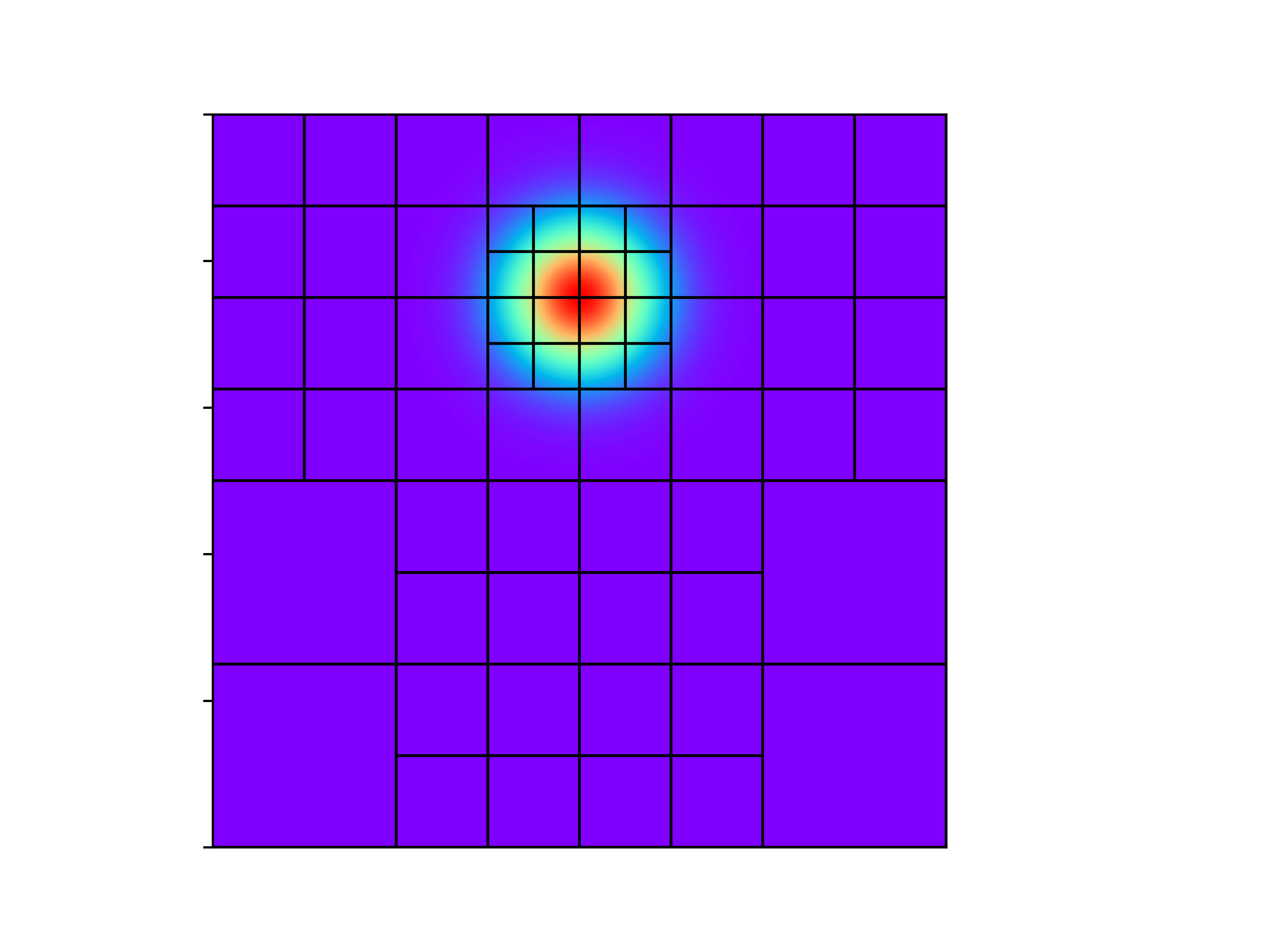}
		\includegraphics[width=.31\textwidth,clip=true,viewport= 60 10 360 310]{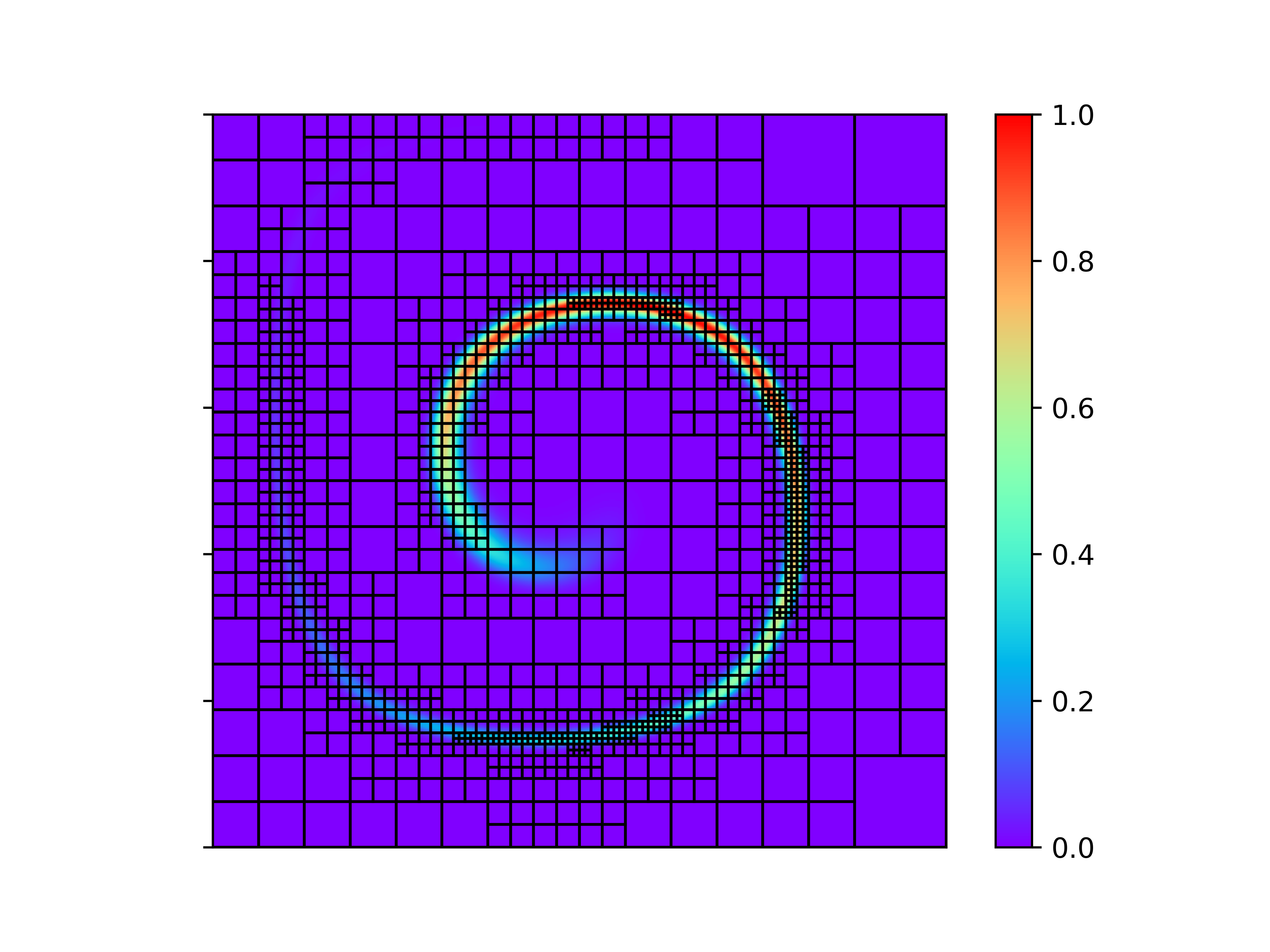} 
		\includegraphics[width=.31\textwidth,clip=true,viewport= 60 10 360 310]{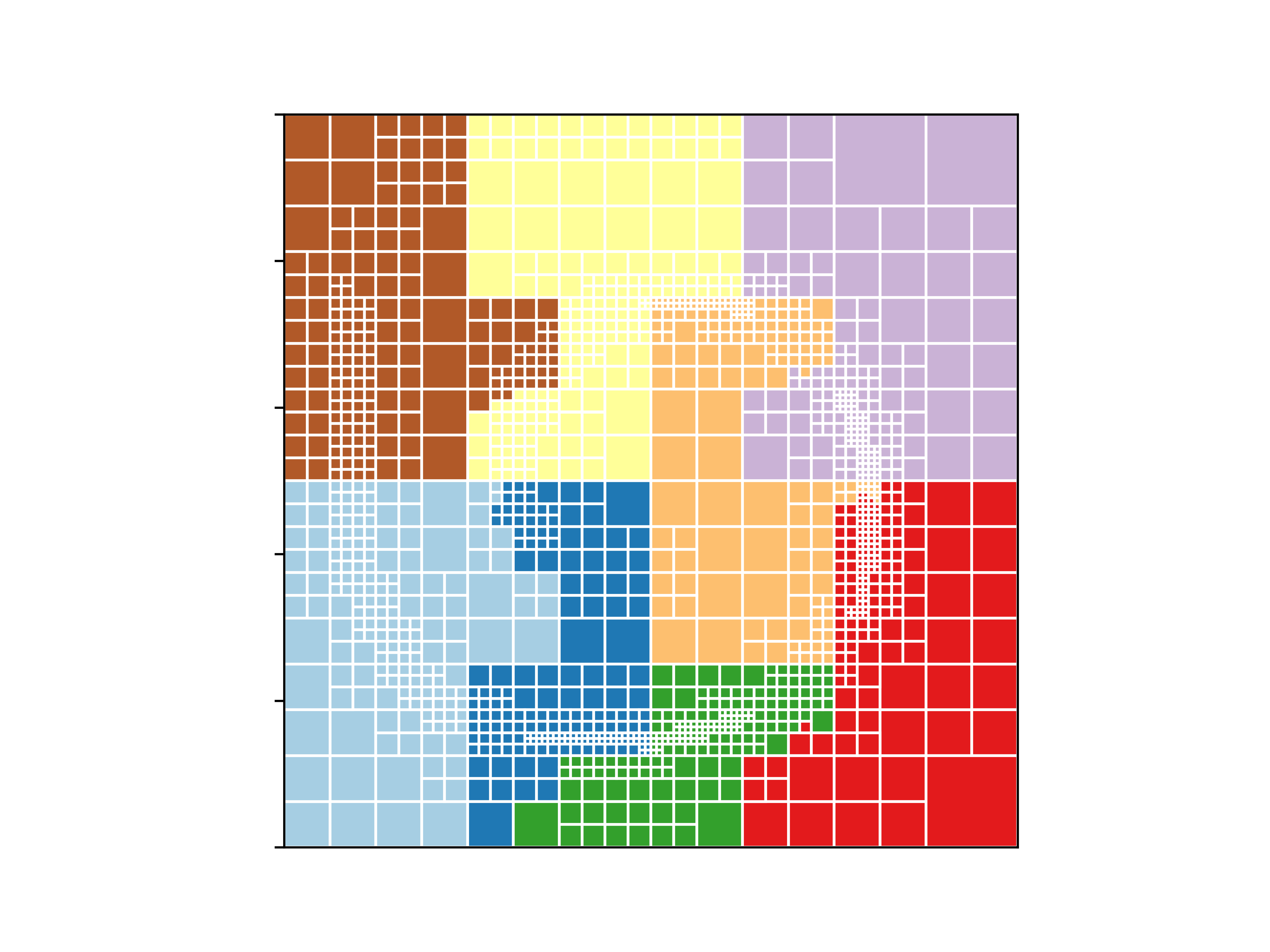}
	\end{center} 
	\vspace{-1em}
	\caption{ 
		Shown is a pseudocolor-plot of $\varphi$ at times $t=0$, $t=t_a/2=2.5 $
			and  the distribution of the blocks among the MPI processes by different colors at $t=2.5$   (from left to right). Each block covers $33\times33$   
		points.}
	\label{fig_swirl_001} 
\end{figure}

Spatial derivatives are discretized with a 4th-order, central finite-difference scheme and we use a 4th-order Runge--Kutta time integration.
Interpolation for the refinement operator is also 4th order. We compute
the solution for $t_{a}=5$, for various maximal  mesh levels $J_{\mathrm{max}}$. 
 The computational domain is a unit square
and we use a block size of $33\times33$.

Figure \ref{fig_swirl_001} illustrates $\varphi$ at the initial
time, $t=0$, and the instant of maximal distortion at $t=2.5 = t_a/2$. 
At $t=2.5$ the grid is strongly refined in
regions of fine structures, while the remaining part of the domain
features a coarser resolution, e.g., in the center of the domain.
Further the distribution among the MPI processes is shown by different colors, revealing the locality of the space filling curve.

In the following  we compare soloutions with the finest strutures at $t=t_a/2$ with a reference solution, to investigate the quality and performance. 
	The reference solution is obtained with a pseudo-spectral code on a sufficiently fine mesh to have a negligible error compared with the current results.

\begin{figure}[htb]
\begin{center}
\includegraphics[width=0.48\textwidth, clip=true,viewport= 0 0 450 320 ]{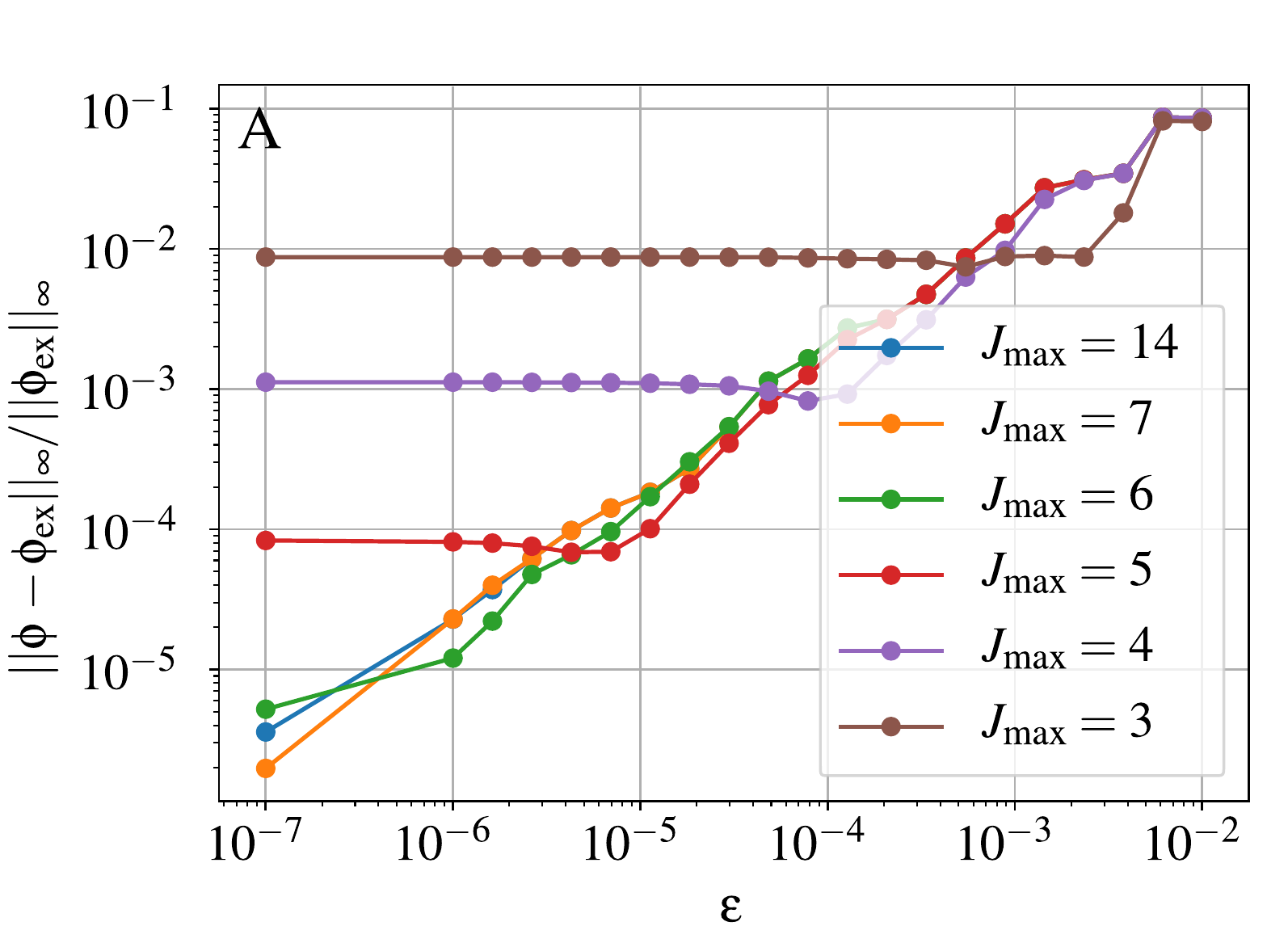}
\includegraphics[width=0.48\textwidth, clip=true,viewport= 0 0 450 320 ]{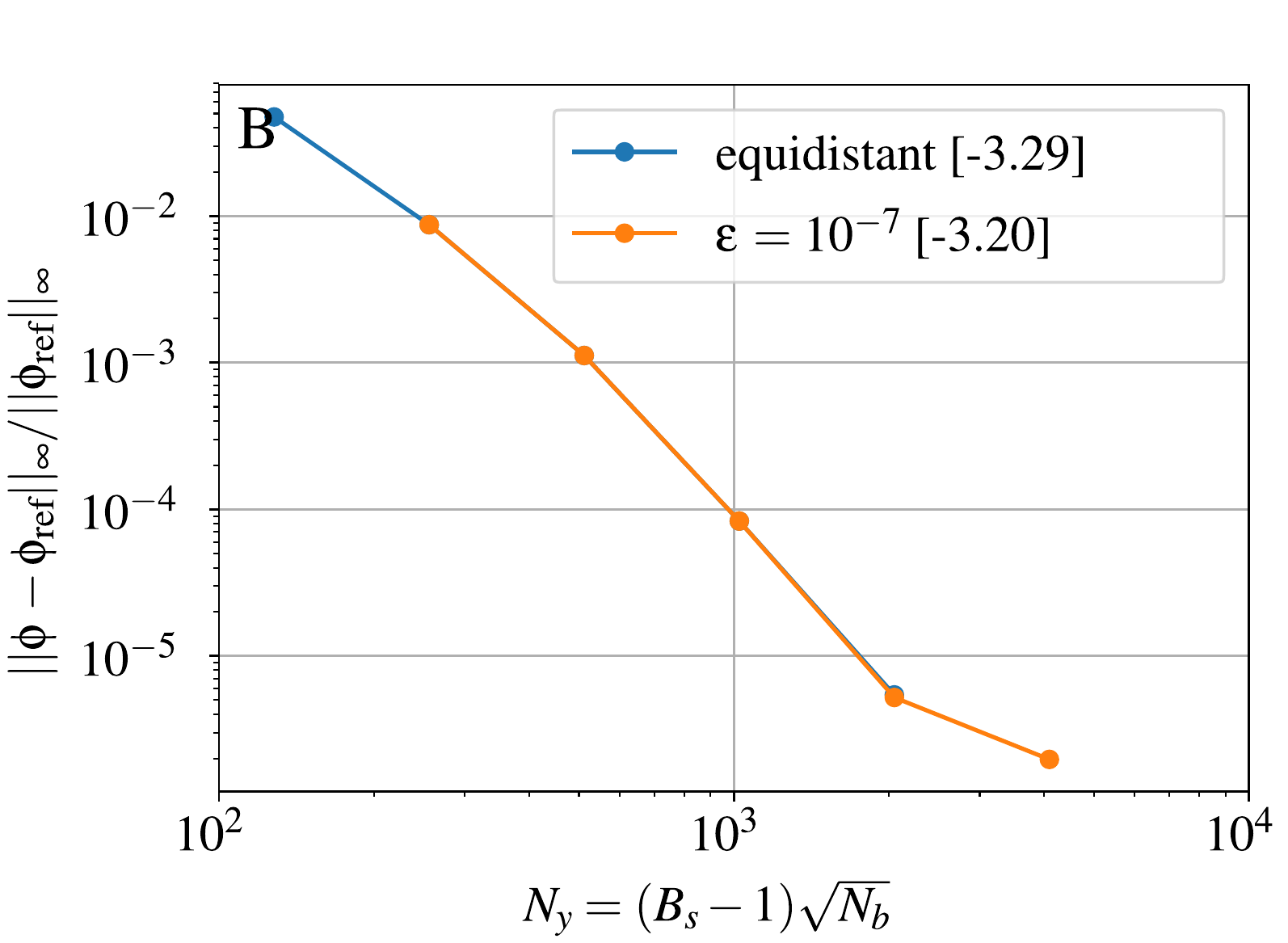}\\[1em]
\includegraphics[width=0.48\textwidth, clip=true,viewport= 0 0 450 320 ]{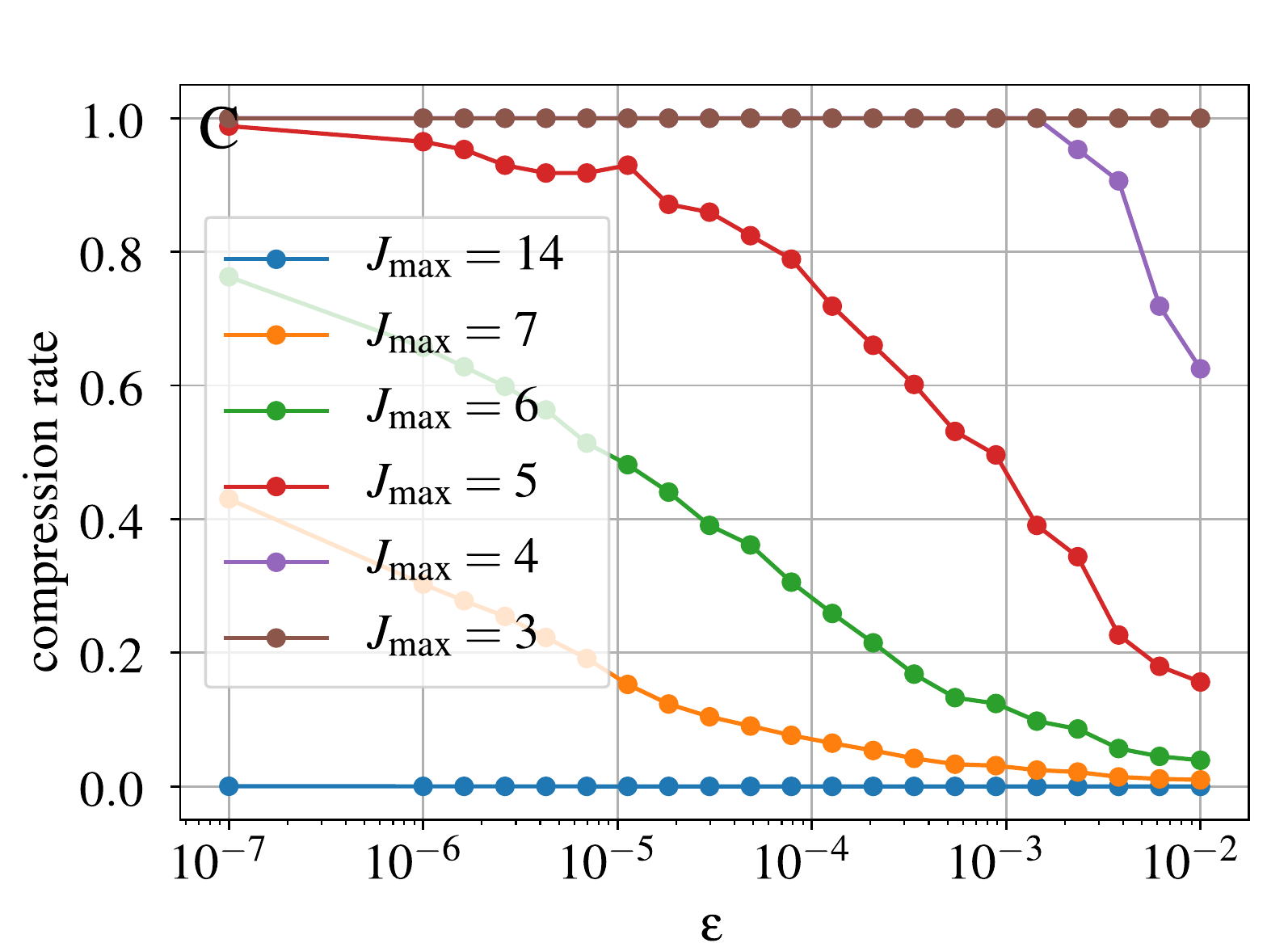}
\includegraphics[width=0.48\textwidth, clip=true,viewport= 0 0 450 320 ]{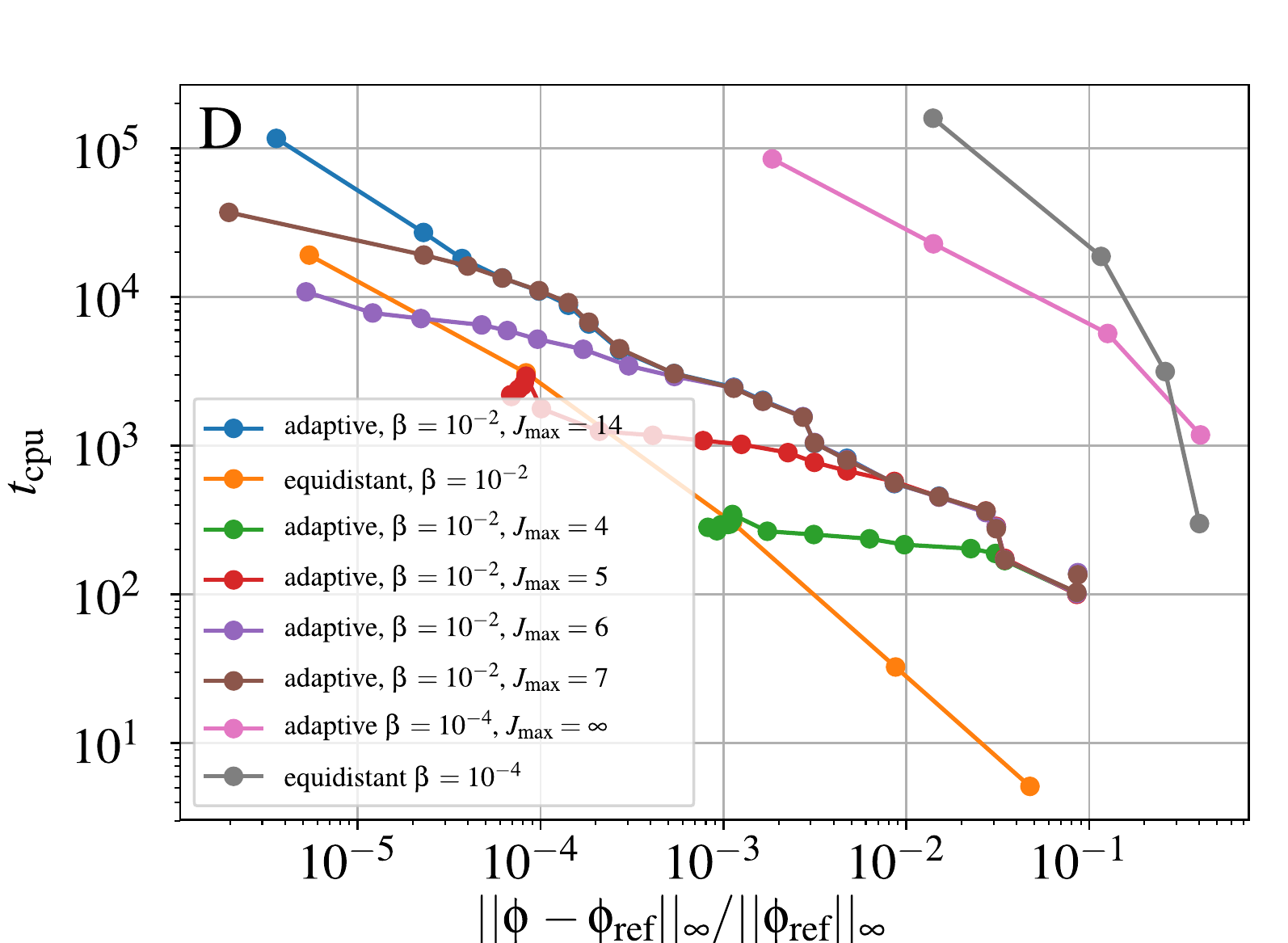}
\end{center}	
	\vspace{-1em}
	\caption{Swirl test for varying $J_{\mathrm{max}}$ and $\varepsilon$.
		A: For different maximal refinement levels a saturation of the error  is seen at different values of $\varepsilon$, showing the cross over form threshold- to discretization-error.
		B: Error decay for fixed   $\varepsilon= 10^{-7}$ and varying $J_{\mathrm{max}}$ (i.e. the rightmost data points in A) as a function of the number of points in one direction. 
		The adaptive computation preserves roughly the 4$^{th}$ order accuracy of the discretization scheme. 
		C: Compression rate defined as block of the adaptive mesh compared with a equidistant mesh constantly  on the same $J_{\mathrm{max}}$. 
		D: The CPU time as a function of discretization error for two different initial conditions. For the broad pulse ($\beta = 10^{-2}$) the adaptive solution is faster for an appropriate choice of $J_{\mathrm{max}}(\varepsilon)$ for the finer pules ( $\beta = 10^{-2}$) it is faster even for a constant $J_{\mathrm{max}}=\infty$ for  relevant errors.               
	}
	\label{fig_swirl_002} 
\end{figure}

Fig. \ref{fig_swirl_002}A illustrates the relative error, computed as the $\infty$-norm of the difference $\varphi-\varphi_{\mathrm{ex}}$, normalized by $||\varphi_{\mathrm{ex}}||_{\infty}$. 
All quantities are evaluated on the terminal grid. 
A linear least squares fit exhibits convergence orders close to one for the large maximal refinements. 
In this case the error decays, as expected, linearly in $\varepsilon$.
For smaller $J_{\mathrm{max}}$ we find a saturation of the error, which is determined by the highest allowed resolution. 
This different levels are plotted  in Fig. \ref{fig_swirl_002}B, where  a convergence order close to four, as expected by the space and time discretization is found. 
Thus, the  points  where the saturation sets in are turnover points form an threshold to and cut-off dominated error. 
For the sake of efficiency one aims to be close to this turnover point where both errors are of similar size. 
In Fig. \ref{fig_swirl_002}C the compression rate, i.e. the number of blocks relative to an equidistant grid constantly on the level of the same $J_{\mathrm{max}}$ is depicted. 
As expected the compression becomes close to one for small $\varepsilon$. 
In  Fig. \ref{fig_swirl_002}D the error is shown as a function of the computational time for two initial conditions, the broad pulse with $\beta= 10^{-2} $  and a narrower one with $\beta= 10^{-4}$. 
For the broad pulse ($\beta= 10^{-2} $) the curves for different $J_{\mathrm{max}}$ are below the equidistant curve only for carefully chosen values of $\varepsilon$. 
This is explained by the wide area of refinement at the final time, see Fig.~\ref{fig_swirl_001}. Here a multi-resolution method cannot win much.
Even for $J_{\mathrm{max}}=14$, which in practice means deactivating the level restriction,  a similar scaling as for the equidistant grid is found with a factor approaching about four. 
Thus, even without tuning $J_\mathrm{max}(\epsilon)$ accordingly, and given the low cost of the right hand side, the computational complexity of the adaptive code scales reasonably compared to the equidistant solution.
%
For a finer initial condition ($\beta= 10^{-4}$), even without the level restriction ( $J_\mathrm{max} = \infty$),  the adaptive code produces better run-times for practical relevant errors.    

\section{Navier-Stokes test case} \label{dsl}
In this section we present the results of a second test case, governed by the ideal-gas, constant heat capacity compressible Navier-Stokes equations in the skew-symmetric formulation \cite{ReissSesterhenn2014}.  
 A double
shear-layer in a periodic domain is perturbed so that the growing
instabilities end up with small scale structures, similar to \cite{MaulikSan2017}.
The size of the computational domain is $L=L_{x}=L_{y}=8$ and the
shear layer is initially located at $\frac{L}{2}\pm0.25$.
The density
and $y$-velocity is $\rho_{1}=2$ and $v_{1}=1$ between the shear
layers and $\rho_{0}=1$ and $v_{0}=-1$ otherwise. At the jumps it
is smoothed  by $\tanh((y-y_\mathrm{jump})/\lambda_{w})$ with a width $\lambda_{w}=L/240$.
The the initial pressure is uniformly $p=2.5$. The $x$-velocity is
disturbed to induce the instability in a controlled manner by $u=\lambda\sin(2\pi(y-L/2))$
with $\lambda=0.1$. The dynamic viscosity is given
by $\mu=10^{-6}$. The adiabatic index is $\gamma=1.4$ and the Prandtl
number is $Pr=0.71$ and the specific gas constant $R_\mathrm{s} = 287.05$. 

We discretize spatial derivatives with standard 4th-order central
differencing scheme, use the standard 4th-order Runge-Kutta time integration
and for interpolation a 4th-order scheme. 
We use global time stepping so that the time step is (usually) determined by the time step at the highest
mesh level. We apply a shock capturing filter as described in \cite{BogeyCacquerayBailly2009} with a
threshold value of $r_\mathrm{th}=10^{-5}$ in every time step. 
	Filtering, as any procedure to suppress high wave numbers (e.g. flux limiter, slope limiter or numerical damping), interacts with the MR.
	No special modification beyond the previously described \cite{ReissSesterhenn2014} smoothed detector, was necessary for the use with the multi resolution framework. 
	The investigation of the interplay between filtering and MR is left for future work.  

\begin{figure}[tb]
	\begin{center}
		\includegraphics[width=.7\textwidth, clip=true, viewport= 0 0 590 590 ]{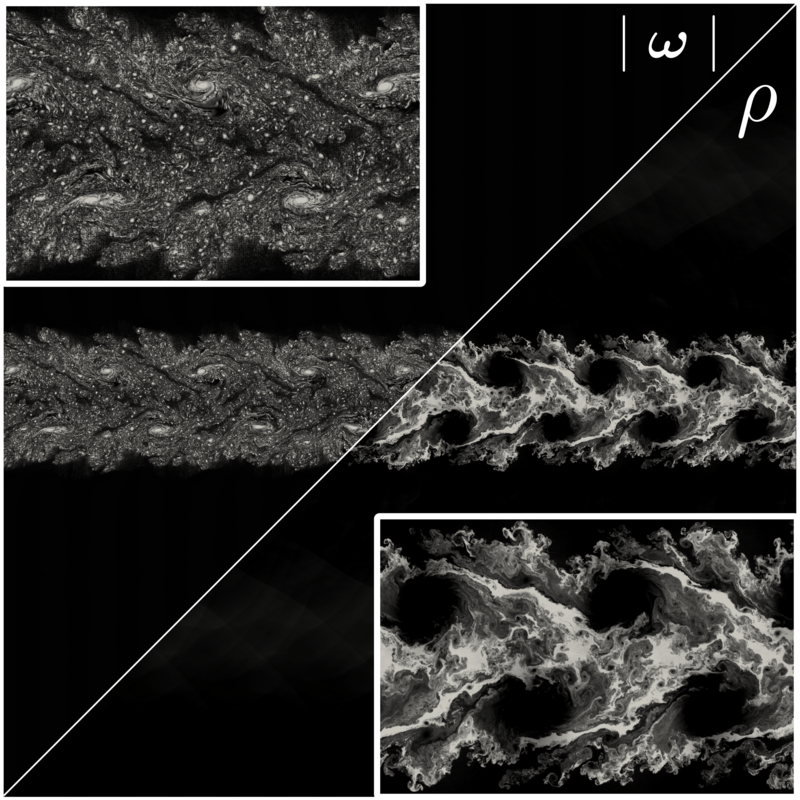}
	\end{center}
	\vspace{-1em}
	\caption{Double shear layer, plot  of density $\rho$  and the absolute value of the vorticity $|\omega|$  at
		time $t=4$ on an adaptive grid with threshold value $\varepsilon=1$e-$3$,
		maximum mesh level  $J_{\mathrm{max}}=8$. 
		\label{fig_dsl002} } 
	
\end{figure}

In Fig. \ref{fig_dsl002} the density field for adaptive computations with a threshold $\varepsilon=10^{-3}$ at  $t=4$ is shown. 
In both density and vorticity field one can observe small scale structures created by the shear layer instability. 
The size and form of the structures are in agreement with  \cite{MaulikSan2017}.





\begin{figure}[htb]
\begin{center}
	\includegraphics[width=0.49\textwidth,clip=true,viewport= 0 180 550 600]{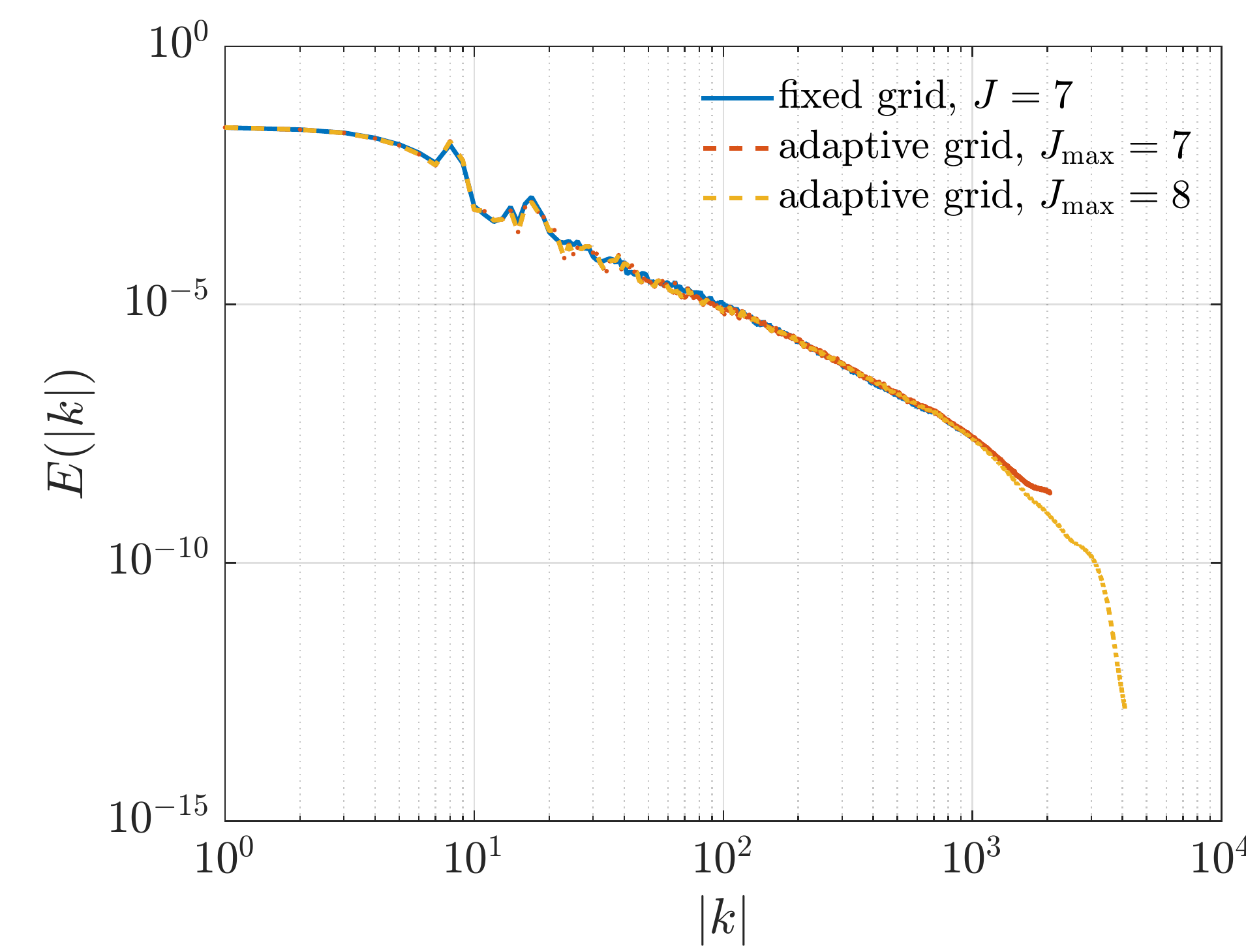} 
	\includegraphics[width=0.49\textwidth,clip=true,viewport= 0 180 550 600]{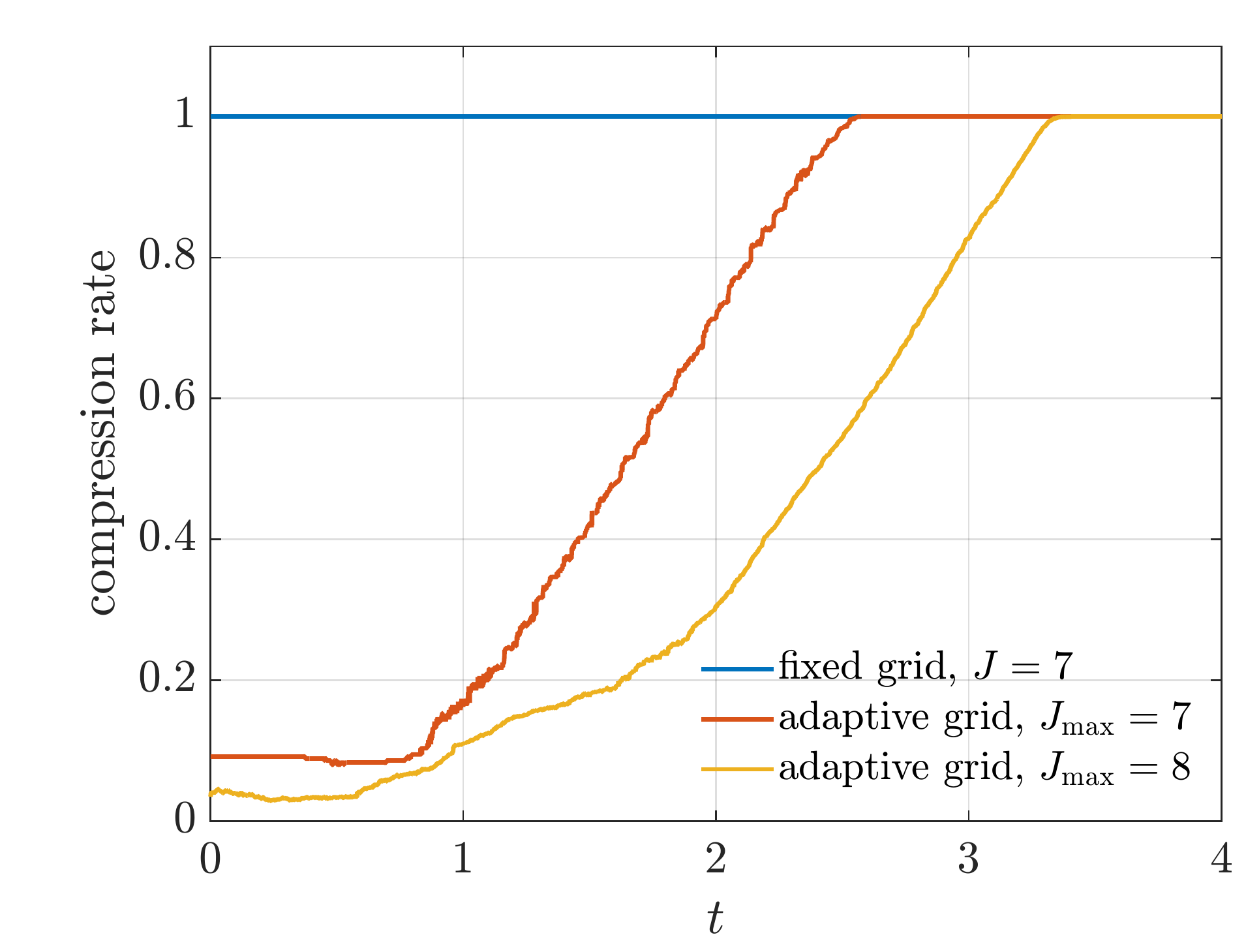} 
\end{center}
	\vspace{-1em}
\caption{Left: Energy spectrum of the double shear layer. The  computations
were performed on a  fixed grid with mesh level $J=7$, and on adaptive grids with threshold
value $\varepsilon=10^{-3}$, maximum mesh level $J_{\mathrm{max}}=7$, $J_{\mathrm{max}}=8$, $t=4$.
Right: The compression rate. After high initial copressions  the grid fills up due to high wavenumber acoustic waves. 
\label{fig_dsl004}}
 
\end{figure}

	In the right of Fig.~\ref{fig_dsl004} the compression rate of the shear layer is plotted over time. 
	We start with a low number of blocks  (i.e. low values of the compression rate),  the grid  fills up to the maximal refinement with simulation time. This is explained by short wavelength acoustic waves emitted by the shear layer. 
	Depending on the investigation target a modified threshold criterion, e.g., applying it only to certain variables might be beneficial. 
	For this the error estimation must be reviewed and it is left for future work.


In Fig. \ref{fig_dsl004} we show the kinetic energy
spectra for these computations compared to the result for a fixed
grid. 
To calculate the energy spectra we refine the mesh after the computation to a fixed mesh level, if needed.
They agree well on the resolved scales. For the higher maximum mesh level $J_{\mathrm{max}}$ we observe a better resolution
	of the small scale structures.  
 Summarized, if we compare
adaptive and fixed mesh computation, we can observe a good resolution
of the small scales within the double shear layer.


\begin{figure}[htb]
	\begin{center}
		\includegraphics[width=.45\textwidth, clip=true , viewport= 30 180 570 590]{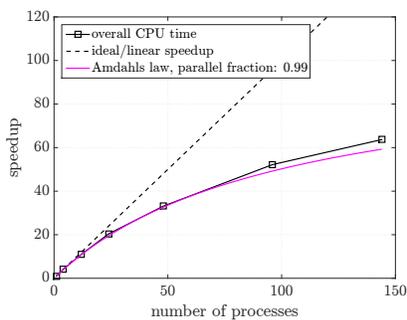} 
	\end{center}
	\vspace{-1em}
	\caption{From the strong scaling a parallel fraction of 99\% can be estimated.  
		\label{fig_dsl003} } 
	
\end{figure}

Fig.~\ref{fig_dsl003} shows the strong scaling behavior for the adaptive double shear layer computation with $J_{\mathrm{max}}$ = 7. 
We observe a scaling which is predicted by Amdahl's law with a parallel fraction of 0.99. 
The observed strong scaling is reasonable and we anticipate that code
 optimization will yield further improvements.

\section{Conclusions and Perspectives} \label{con}

The novel framework \verb+WABBIT+ with its main structures and concepts
has been described. \verb+WABBIT+ uses a multiresolution algorithm
to adapt the mesh to capture small localized structures. Within the
framework different equation sets can be used. 

We showed that the error due to the thresholding is controlled and
scales nearly linear. In the Gaussian pulse test case we found  that the maximum number of blocks is reached at the largest
deformation of the pulse and after that the mesh is coarsen with several
orders of magnitude. 
We observed that the fill-up was strongly reduced by using a symmetric
interpolation stencil, which will be investigated in future work.

In the second test case we showed an application of the compressible
Navier-Stokes equations. Here we saw a good resolution of small scale
structures and observed the impact of discarding wavelet coefficient
on the physics of the shear layer. 
In our simulations we observe a reasonable strong scaling. 
Scaling	will be assessed in more detail when foreseen improvements are implemented.

In the near future we will extend the physical situation by using
reactive Navier-Stokes equations to simulate turbulent flames. 
Validation for 3D problems and further improvement of the performance
is currently worked on. For this an additional parallelization with
\verb|openMP| is in preparation, which should reduce the communication
effort further in typical cluster architecture. Further a generic
boundary handling within the frame work and an interface to connect
other \verb+MPI+ programs is under way.

\paragraph{Acknowledgments.}

MS and JR thankfully acknowledge funding by the Deutsche Forschungsgemeinschaft
(DFG) (grant SFB-1029, project A4). TE and KS acknowledge financial
support from the Agence nationale de la recherche (ANR Grant 15-CE40-0019)
and DFG (Grant SE 824/26-1), project AIFIT. This work was granted
access to the HPC resources of IDRIS under the allocation 2018-91664
attributed by GENCI (Grand \'Equipement National de Calcul Intensif).
For this work we were also granted access to the HPC resources of
Aix-Marseille Universit\'e financed by the project Equip@Meso (ANR-10-EQPX-
29-01). TE and KS thankfully acknowledge financial support granted
by the minist\`eres des Affaires \'etrang\`eres et du d\'eveloppement International
(MAEDI) et de l\textquoteright Education national et l'enseignement
sup\'erieur, de la recherche et de l'innovation (MENESRI), and the Deutscher
Akademischer Austauschdienst (DAAD) within the French-German Procope
project FIFIT.




%
%
\bibliography{local_wabbit}{}
\bibliographystyle{plain}

\end{document}